\input phyzzx.tex
\tolerance=1000
\voffset=-0.0cm
\hoffset=0.7cm
\sequentialequations
\def\rl{\rightline}

\def\t1{{\tilde 1}}

\def\PLB[arXiv:#1#2#3{Phys. Lett. B {\bf#1} (19#2) #3}

\def\t{\theta}

\REF{\SUSY}{Y. Shadmi and Y. Shirman, Rev. Mod. Phys. {\bf 72} (2000) 25, [arXiv:hep-th/9907225]; J. Terning, [arXiv:hep-th/0306119]; K. Intriligator and N. Seiberg, 
Class. Quant. Grav. {\bf 24} (2007) S741, [arXiv:hep-th/0702069].}
\REF{\ISS}{K. Intriligator, N. Seiberg and D. Shih, JHEP {\bf 0604} (2006) 021, [arXiv:hep-th/0602239].}
\REF{\RAY}{S. Ray, Phys. Lett {\bf B642} (2006) 13, [arXiv:hep-th/0607172].}
\REF{\KIT}{R. Kitano, [arXiv:hep-ph/0606129]; Phys. Lett. {\bf B641} (2006) 203, [arXiv:hep-ph/0607090].}
\REF{\DM}{M. Dine and J. Mason, Phys. Rev. {\bf D77} (2008) 016005, [arXiv:hep-ph/0608063].}
\REF{\DIN}{M. Dine, J. L. Feng and E. Silverstein, Phys. Rev. {\bf D74} (2006) 095012, [arXiv:hep-th/0608159].}
\REF{\KOO}{R. Kitano, H. Ooguri and Y. Ookouchi, Phys. Rev. {\bf D75} (2007) 045022, [arXiv:hep-ph/0612139].}
\REF{\MUR}{H. Murayama and Y. Nomura, Phys. Rev. Lett. {\bf 98} (2007) 151803, [arXiv:hep-ph/0612186].}
\REF{\OS}{O. Aharony and N. Seiberg, JHEP {\bf 0702} (2007) 054, [arXiv:hep-ph/0612308].}
\REF{\CST}{C. Csaki, Y. Shirman and J. Terning, JHEP {\bf 0702} (2007) 099, [arXiv:hep-ph/0612241].}
\REF{\SHI}{K. Intriligator, N. Seiberg and D. Shih, JHEP {\bf 0707} (2007) 017, [arXiv:hep-th/0703281].}
\REF{\HAB}{N. Haba and N. Mara, Phys. Rev. {\bf D76} (2007) 115019, arXiv:0709.2945[hep-ph].}
\REF{\KD}{K. R. Dienes and B. Thomas, Phys. Rev. {\bf D78} (2008) 106011, arXiv:0806.3364[hep-th]; Phys. Rev. {\bf D79} (2009) 045001, arXiv:0811.3335[hep-th].}
\REF{\GIV}{A. Giveon and D. Kutasov, Nucl. Phys. {\bf B796} (2008) 25, arXiv:0710.0894[hep-th].}
\REF{\ZKKS}{A. Giveon, A. Katz, Z. Komargodski and D. Shih, JHEP {\bf 0810} (2008) arXiv:0808.2901[hep-th].}
\REF{\KOM}{Z. Komargodski and D. Shih, JHEP {\bf 0904} (2009) 093, arXiv:0902.0030[hep-th].}
\REF{\BGH}{I. Bena, E. Gorbatov, S. Hellerman, N. Seiberg and D. Shih, JHEP{\bf 0611} (2006) 088, [arXiv:hep-th/0608157].}
\REF{\FRA}{S. Franco, I. Garcia-Etxebarria and A.M. Uranga, JHEP {\bf 0701} (2007) 085, [arXiv:hep-th/0607218].}
\REF{\AMI}{A. Giveon and D. Kutasov, Nucl. Phys. {\bf B778} (2007) 129, [arXiv:hep-th/0703135]; JHEP {\bf 0802} (2008) 038, arXiv:0710.1833[hep-th].}
\REF{\ARG}{R. Argurio, M. Bertolini, S. Franco and S. Kachru, JHEP {\bf 0706} (2007) 017, [arXiv:hep-th/0703236].}
\REF{\BER}{M. Bertolini, S. Franco and S. Kachru, JHEP {\bf 0701} (2007) 017, [arXiv:hep-th/0610212].}
\REF{\OOG}{H. Ooguri and Y. Ookouchi, Phys. Lett. {\bf B641} (2006) 323, [arXiv:hep-th/0607183].}
\REF{\AGA}{M. Aganagic, C. Beem, J. Seo and C. Vafa, [arXiv:hep-th/0610249].}
\REF{\ABK}{M. Aganagic, C. Beem ad S. Kachru, Nucl. Phys. {\bf B796} (2008) 1, arXiv:0709.4277[hep-th].}
\REF{\OKS}{O. Aharony, S. Kachru and E. Silverstein, Phys. Rev. {\bf D76} (2007) 126009, arXiv:0708.0493[hep-th].}
\REF{\AMA}{A. Amariti, D. Forcella, L. Girardello and A. Mariotti, JHEP {\bf 0812} (2008) 079, arXiv:0803.0514[hep-th].}
\REF{\FU}{S. Franco and A. M. Uranga, JHEP {\bf 0606} (2006) 031, [arXiv:hep-th/0604136].}
\REF{\AHN}{C. Ahn, Class. Quant. Grav. {\bf 24} (2007) 1359, [arXiv:hep-th/0608160]; Class. Quant. Grav. {\bf 24} (2007) 3603, [arXiv:hep-th/0702038].}
\REF{\TAT}{R. Tatar and B. Wettenhall, Phys. Rev. {\bf D76} (2007) 126011, arXiv:0707.2712[hep-th].}
\REF{\VER}{M. Buican, D. Malyshev and H. Verlinde, JHEP {\bf 0806} (2008) 108, arXiv:0710.5519[hep-th].}
\REF{\LAST}{E. Halyo, arXiv:0906.2127[hep-ph].} 
\REF{\INS}{R. Blumenhagen, M. Cvetic and T. Weigand, Nucl. Phys. {\bf B771} (2007) 113, [arXiv:hep-th/0609191]; L. Ibanez and A. Unranga JHEP {\bf 0703} (2007) 052, 
[arXiv:hep-th/0609213]; B. Florea, S. Kachru, J. McGreevy and N. Saulina, JHEP {\bf 0705} (2007) 024, [arXiv:hep-th/0610003];
R. Blumenhagen, M. Cvetic and S. Kachru, T. Weigand, arXiv:0902.3251[hep-th] and references therein.}
\REF{\GK}{A. Giveon and D. Kutasov, Rev. Mod. Phys. {\bf 71} (1999) 983, [arXiv:hep-th/9802067].}
\REF{\SING}{F. Cachazo, K. Intriligator and C. Vafa, Nucl. Phys. {\bf B603} (2001), [arXiv:hep-th/0103067].}
\REF{\SUP}{E. Witten, Nucl. Phys. {\bf B507} (1997) 658, [arXiv:hep-th/9706109]; M. Aganagic and C. Vafa, [arXiv:hep-th/0012041].}
\REF{\DOUG}{M. Douglas, JHEP {\bf 9707} (1997) 004, [arXiv:hep-th/9612126].}
\REF{\GEO}{C. Vafa, Journ. Math. Phys. {\bf 42} (2001) 2798, [arXiv:hep-th/0008142]; F. Cachazo, K. Intriligator and C. Vafa, Nucl. Phys. {\bf B603} (2001) 3, [arXiv:hep-th/0103067].}
\REF{\NPS}{S. Gukov, C. Vafa and E. Witten, Nucl. Phys. {\bf B584} (2000) 69, Erratum-ibib, {\bf B608} (2001) 477, [arXiv:hep-th/9906070].}
\REF{\DINF}{E. Halyo, Phys. Lett. {\bf B387}  (1996) 43, [arXiv:hep-ph/9606423]; P. Binetruy and G. Dvali, Phys. Lett. {\bf B388}  (1996) 241, [arXiv:hep-ph/9606342].}
\REF{\EDI}{E. Halyo, JHEP {\bf 0407} (2004) 080, [arXiv:hep-th/0312042]; [arXiv:hep-th/0402155]; [arXiv:hep-th/0405269].}

\singlespace
\rl{SU-ITP-09/28}
\pagenumber=0
\normalspace
\medskip
\bigskip
\titlestyle{\bf{Metastable Supersymmetry Breaking Vacua in Abelian Brane Models}}
\smallskip
\author{ Edi Halyo{\footnote*{e--mail address: halyo@stanford.edu}}}
\smallskip
\centerline {Department of Physics} 
\centerline{Stanford University} 
\centerline {Stanford, CA 94305}
\smallskip
\vskip 2 cm
\titlestyle{\bf ABSTRACT}

We construct Abelian brane models with metastable vacua which are obtained from deformations of ${\cal N}=2$ supersymmetric brane configurations.
One such model lives on a D4 brane stretched between two displaced and rotated NS5 branes. Another one lives on a D5 brane wrapped on a deformed and fibered 
$A_2$ singularity.

\singlespace
\vskip 0.5cm
\endpage
\normalspace

\centerline{\bf 1. Introduction}
\medskip

A realistic descripion of physics beyond the Standard Model requires understanding the mechanism of supersymmetry breaking and the phenomenology derived from it.
Until recently, models of supersymmetry breaking were assumed to have a unique, stable and supersymmetry breaking vacuum[\SUSY]. However, it was realized that supersymmetry may also be broken 
in a metastable vacuum of a model which preserves supersymmetry[\ISS]. After ref. [\ISS], many models with this property have been built in field theory[\RAY-\KOM]. 
Naturally, it is important to build such models not only in field theory but also in string theory. As a result, there has been considerable progress in constructing D--brane models with metastable
vacua[\BGH-\VER]. 

In this letter, we build brane models with metastable vacua which break supersymmetry at tree level. These are based on recently built models in field theory[\LAST] which are basically
deformed ${\cal N}=2$ supersymmetric models. These models have an Abelian gauge symmetry and a matter sector that consists of
two charged and one neutral field coupled through a Yukawa coupling. In addition, the superpotential contains the deformations given by 
mass terms for the charged and singlet fields, singlet F and/or anomalous D--terms. As is well--known, the singlet mass breaks supersymmetry from ${\cal N}=2$ to ${\cal N}=1$. 
Even though these models do not break supersymmetry dynamically they are still interesting. (Dynamical supersymmetry breaking may be achieved by retrofitting[\DIN] in field theory and 
through D--brane instantons in string theory.[\ABK,\INS]) 
First, since they are obtained through different deformations of ${\cal N}=2$ supersymmetric models they can be easily realized in many brane constructions.
In this letter, we construct two such D--brane models: the first is based on intersecting branes[\GK] while the second one is based on branes wrapped on singularities[\SING].
Second, these models can easily be embedded in larger ones with non--Abelian gauge symmetries and a number of fields in the fundamental and adjoint representations.   

The intersecting brane model we construct below describes the theory living on a D4 brane stretched between two NS5 branes. The deformations of the model are given by different 
displacements and rotations of the D4 and NS5 branes[\GK]. In the alternative scenario, the theory lives on two D5 branes wrapped on a deformed $A_2$ singularity fibered on $C(x)$[\SING]. 
The branes wrap the two singularities of $A_2$ that are resolved and/or deformed. In this case, the deformations
of the model are given by the twisting of $A_2$, the location of the singularities on $C(x)$ and of their resolutions. 

This letter is organized as follows. In the next section, we build models with metastable nonsupersymmetric vacua by using intersecting branes. We discuss the cases with F and D--term 
supersymmetry breaking separately. In section 3, we do the same by using
branes wrapped on singularities. Section 4 includes our conclusions and a discussion of our results.

\bigskip
\centerline{\bf 2. Intersecting Brane Models with Metastable Vacua}
\medskip

In this section we describe the intersecting brane model that gives rise to a metastable nonsupersymmetric vacuum at the origin of the field space.
We divide space into $R^{3,1}=(x_0,x_1,x_2,x_3)$ which constitutes our four dimensions and the tranverse space denoted by $x_6$ and
$$v=x_4+ix_5, \qquad w=x_8+ix_9, \qquad  y=x_7 \eqno(1)$$
We consider two parallel NS5 branes along the ($R^{3,1},v$)  directions and separated by a distance $L$ along the $x_6$ direction so that they are at $x_6=0$ and $x_6=L$ (the left and right NS5 branes
respectively). We add a D4 brane stretched between them along the ($R^{3,1},x_6$) 
directions. In addition, we introduce a D6 brane along the ($R^{3,1},y,w$) directions to
the left of the left NS5 brane with a second D4 brane stretched between the D6 and NS5 branes along ($R^{3,1},x_6$) directions. This is a very well understood brane configuration with ${\cal N}=2$
supersymmetry[\GK]. The D4 brane the world--volume theory contains a $U(1)$ gauge field and a neutral scalar  
($\phi$) that parametrizes the location of the D4 brane along the $v$ direction (both arising from the D4--D4 strings).
The second D4 brane (between the D6 and NS5 branes) is frozen due to the orientation of the D6 brane which is perpendicular to the NS5 brane and does not give rise to a gauge 
boson or a singlet. The strings stretched between the two D4 branes give rise to two oppositely charged ($\pm 1$) fields $q_1,q_2$ with the superpotential
$$W=\lambda \phi q_1 q_2 \eqno(2)$$
where $\lambda=g$ is the gauge coupling given by $g^2=g_s \ell_s/L$. 

Now we deform this configuration in a number of ways that changes the superpotential[\GK]. First, we rotate the left NS5 brane 
in the $v-w$ plane by an angle $\alpha$ so that it extends along the $vcos \alpha+ wsin \alpha$ direction. Since the singlet $\phi$ parametrizes the 
distance along the NS5 brane the Yukawa coupling becomes $\lambda=g cos \alpha$. The definition of $\lambda$ involves a subtlety related to the normalization of $\phi$. The previous expression for the Yukawa
coupling is for $\phi$ which is not canonically normalized. If we use a canonical normalization for $\phi$, we need to rescale it by $g$ and then the Yukawa coupling becomes
$\lambda=cos \alpha$.
Second, we add a mass term for the
charged fields $W_1=m q_1 q_2$. This is achieved by moving the D6 brane to $v=-v_0$ which also moves the second D4 brane to that location. Then the two D4 branes are separated and the strings 
stretched between them 
(i.e. $q_1,q_2$) get a mass
$$m={v_0 \over {2 \pi \ell_s^2}} \eqno(3)$$
Third, we add a singlet mass term $W_2=M \phi^2$ by rotating the right NS5 brane by an angle $\alpha+\theta$ in the $v-w$ plane. The relative angle between the NS5 branes gives rise to the singlet mass
$$M={tan \theta \over {2 \pi \ell_s}} \eqno(4)$$
Finally, we add an anomalous D--term by moving the right NS5 brane to a nonzero value of $y=y_0$ leaving the left NS5 brane at $y=0$. The anomalous D--term is given by
$$\xi={y_0 \over {2 \pi g_s \ell_s^3}} \eqno(5)$$
The total superpotential after these deformations becomes 
$$W_{tot}=\lambda \phi q_1 q_2+ m q_1 q_2+ M \phi^2 \eqno(6)$$
The F--terms obtained from $W_{tot}$ are
$$F_{q_1}=(\lambda \phi+m)q_2 \eqno(7)$$
$$F_{q_2}=(\lambda \phi+m)q_1 \eqno(8)$$
and
$$F_{\phi}= \lambda q_1 q_2+ 2M \phi \eqno(9)$$
In addition there is the D--term
$$D=(|q_1|^2-|q_2|^2+ \xi) \eqno(10)$$
The complete scalar potential is given by
$$V=|F_{q_1}|^2+|F_{q_2}|^2+|F_{\phi}|^2+g^2|D|^2 \eqno(11)$$

This potential has a metastable nonsupersymmetric vacuum at the origin of the field space and a supersymmetric vacuum away from the origin. By minimizing the scalar potential in eq. (11), 
it is easy to see that
the origin $\phi=q_1=q_2=0$ is a locally stable vacuum if $m_{q_2}^2=m^2-2g^2 \xi>0$ (since the other masses squared are always positive). 
Supersymmetry is broken at the origin because $D=\xi \not = 0$. All three deformations above are necessary for the origin to be a nonsupersymmetric metastable vacuum. If the singlet 
mass is zero, $M=0$, $\phi$ 
classically becomes a flat direction which is lifted by one--loop effects. The one--loop potential obtained after integrating out the massive charged fields pushes
$\phi$ to the origin. Clearly, if the charged fields have no mass term, $m=0$, then the origin is not stable because
$m_{q_2}^2=-2g^2 \xi<0$ and $q_2$ is tachyonic. If the anomalous D--term is zero, $\xi=0$, then the origin is stable but also supersymmetric.

The supersymmetric vacua are are given by
$$\phi=-{m \over \lambda} \qquad q_1={{2Mm} \over {\lambda^2 q_2}} \eqno(12)$$
where
$$|q_2|^2={\xi \over 2} \pm {1 \over {2 \lambda^2}}\sqrt{\lambda^4 \xi^2+16 M^2m^2} \eqno(13)$$
Clearly only the plus sign above makes sense since $|q_1|^2$ cannot be negative.
Eq. (13) fixes the VEV of $q_1$ up to a phase and therefore the space of supersymmetric vacua is one dimensional (which is a circle of radius $|q_1|$ parametrized by the phase). Note 
that this is not the moduli space because the
superpotential and the scalar potential are independent of this phase. We can pick any phase for the VEV of $q_1$ and for simplicity we set it to zero, i.e take $q_1$ to be real. Then
we get two real values for $q_1$ corresponding to the two supersymmetric vacua. 
We note that all three VEVs for $\phi,q_1,q_2$ are inversely proportional to $\lambda$ so the supersymmetric vacuum can be made arbitrarily far from the origin for small enough $\lambda$.
This results in a 
long--lived, metastable vacuum at the origin. As expected, when the Yukawa coupling vanishes, $\lambda \to 0$, the supersymmetric vacuum runs to
infinity and disappears. 

Let us now describe the two vacua in terms of the brane construction.
The metastable brane configuration was described above. In it the two D4 branes are separated by $v_0$ in the $v$ direction but are at the same point in the remaining four directions, i.e.
both are at $w=y=0$ with no Wilson line along $x_6$. The left and right NS5 branes are rotated by angles $\alpha$ 
and $\alpha+\theta$ respectively. In addition the right NS5 is displaced by $y_0$ in the $y$ direction. This configuration of branes is metastable, i.e. a small perturbation of
of the D4 brane will return it to the original configuration. 
The supersymmetric vacuum is given by eqs. (12) and (13). In this configuration, the two D4 branes are both at $v=-v_0$ since $\phi$ parametrizes the $v$ coordinate of the D4 brane
$$\phi={v \over {2 \pi \ell_s^2}} \eqno(14)$$
and separated along $w,y$ by (assuming a vanishing Wilson line along $x_6$)
$$q_1={w \over {2 \pi \ell_s^2}} \qquad q_2={y \over {2 \pi \ell_s^2}} \eqno(15)$$
the VEVs of $q_1,q_2$. 
In fact the two D4 branes are connected forming one long D4 brane stretched from the right NS5 brane to the D6 brane bypassing the left NS5 brane at $y=0$.
We can also see from the brane configuration that as $\lambda \to 0$ the supersymmetric vacuum escapes to infinity. As $\lambda$ decreases, the left NS5 brane turns towards the $w$
direction, i.e. the D6 brane. As this happens, the D4 brane has to go further along the left NS5 brane (the $v$ direction) in order to reach the point $-v_0$ where it meets the second
D4 brane. In fact, from the geometry of the brane configuration it is easy to see that the displacement has to be $-v_0/cos \alpha$ which agrees with the VEV of $\phi$.
When $\lambda=0$ the two D4 branes can never meet, i.e. the supersymmetric vacuum escapes to infinity.

Above, supersymmetry was broken by a nonzero D--term in the metastable vacuum at the origin of field space. We can just as easily construct a model with F--term supersymmetry breaking[\LAST]. 
This requires a vanishing anomalous D--term, $\xi=0$ in addition to a nonzero singlet F--term, $F\phi$, in the superpotential. The nonzero F--term can be obtained by moving the D4 brane 
(between the NS5 branes) to $w_0$ in the $w$ direction (rather than in the $y$ direction which gives rise to a D--term) so that 
$$F={w_0 \over {2 \pi g_s \ell_s^3}} \eqno(16)$$
The rotations of the NS5 branes remain as before i.e. the left and right NS5 branes are rotated in the
$v-w$ plane by angles $\alpha$ and $\alpha+\theta$ respectively. Now the superpotential becomes
$$W=\lambda \phi q_1 q_2+ mq_1 q_2+ M \phi^2 +F \phi \eqno(17)$$
where $\lambda$ and $m$ are defined as before. The F--terms for $q_1$ and $q_2$ are still given by eqs. (7) and (8). The F--term for the singlet becomes
$$F_{\phi}=\lambda q_1 q_2+2M \phi+ F \eqno(18)$$
The scalar potential is still given by eq. (11) using eqs. (7),(8) and (18) where the D--term contribution is given by eq.(10) with $\xi=0$.

This model was investigated in field theory in ref. [\LAST] where it was found that a complete analysis of the vacuum structure is somewhat complicated. Therefore, 
for simplicity and since our aim is only to demonstrate the existence of a metastable nonsupersymmetric vacuum, we assume that the D--flatness condition is satisfied by taking $q_1=q_2$
whereas the most general condition is $|q_1|=|q_2|$.
In addition, we consider only real VEVs for all fields. With these simplifications, there is a supersymmetric vacuum at
$q_1=q_2=0$ and $\phi=-F/2M$. (There is another supersymmetric vacuum at $\phi=-m/\lambda$ and $q_1=q_2=\sqrt{2Mm/\lambda^2-F/\lambda}$.)
Note that this vacuum is not parametrically far from the origin of field space and in fact two of the VEVs vanish. (This should be contrasted with the second supersymmetric vacuum 
which is parametrically far from the origin.) This vacuum corresponds to a brane configuration in which both D4 branes are at $w=y=0$; however the right one is at 
$v=-F \pi \ell_s^2/ M$ whereas the left one is at $-v_0$ as above.

In this case, the metastable nonsupersymmetric vacua are at[\LAST] 
$$q_1=q_2=\pm \sqrt{{M(\lambda \phi+m)} \over \lambda^2} \eqno(19)$$
where the singlet VEV is given by
$$\phi={1 \over {2 \lambda^2}}[-(3 \lambda M+2 \lambda m) \pm \sqrt{(3 \lambda M+ 2\lambda m)^2-4 \lambda^2(m^2+\lambda F+mM)}] \eqno(20)$$
Clearly, the above vacua break supersymmetry since all three F--terms are nonzero there.
In particular corners of the parameter space, it is easy to show that the above vacua are metastable[\LAST]. For example, in the limit $m>>M>>\sqrt{F}$ ($\sqrt{|F|}>>M>>m$)
the vacua are locally stable for $M>0$ ($16M^2>\lambda |F|$ with $F<0$.) Moreover, since the VEVs in eqs. (19) and (20) are inversely proportional to
$\lambda$, the nonsupersymmetric vacua are parametrically far from the supersymmetric ones. Therefore, they can be metastable for a small enough $\lambda$.



\bigskip
\centerline{\bf 3. Metastable Vacua from Branes on Singularities}
\medskip

In this section, we construct models with metastable supersymmetry breaking vacua by using D5 branes wrapped on deformed $A_2$ singularities fibered over the complex plane $C(x)$. 
Consider the deformed $A_2$ singularity given by
$$uv=(z-3mx)(z-mx)(z+m(x-2a)) \eqno(21)$$
where $x$ is the complex coordinate of $C(x)$. The zeros of each factor denoted by $z_i$, $i=1,2,3$ are related to the holomorphic volume of the two $S^2$s, $\alpha_{1,2}$, that 
deform the $A_2$ singularity
$$\alpha_1=z_1-z_2, \qquad \alpha_2=z_2-z_3 \eqno(22)$$
Now we wrap one D5 brane on each $S^2$ above. The $3+1$ dimensional world--volume theory[\SING] is an ${\cal N}=1$ supersymmetric $U(1)_1 \times U(1)_2$ gauge theory with
two singlets ($\Phi_1,\Phi_2$) that parametrize the locations of the two D5 branes on the complex plane $C(x)$. In addition, there are two oppositely charged fields ($Q_{12},Q_{21}$)
that arise from the strings stretched between the two D5 branes. The superpotential of the model is[\SING]
$$W_1=Q_{12}\Phi_2 Q_{21}-Q_{12}\Phi_1 Q_{21} \eqno(23)$$
Fibering the singularity over $C(x)$ results in a superpotential for the singlets which is given by[\SUP]
$$W_2=\int^{\Phi_1} \alpha_1 dx +\int^{\Phi_2} \alpha_2 dx \eqno(24)$$
Using the definition of $\alpha_i$ in eq. (22) and $z_1=3mx$, $z_2=mx$ and $z_3=-m(x-2a)$ from eq. (21) we obtain the singlet superpotential
$$W_2=m\Phi_1^2+ m(\Phi_2-a)^2 \eqno(25)$$
so that the total superpotential of the model becomes
$$W=W_1+W_2=Q_{12} Q_{21}(\Phi_2-\Phi_1)+m\Phi_1^2+ m(\Phi_2-a)^2 \eqno(26)$$
We can also include an anomalous D--term for $U(1)_1-U(1)_2$ under which $Q_{12}$ and $Q_{21}$ are charged (the orthogonal combination $U(1)_1+U(1)_2$ describes the center of mass motion and decouples
from matter)
$$D=|Q_{12}|^2-|Q_{21}|^2+\xi \eqno(27)$$
The anomalous D--term is related to the small resolution (Kahler deformation) of the singularity by blowing up two $S^2$s[\DOUG] 
$$4 \pi g_s \xi=\int_{S_1^2} J -\int_{S_2^2} J \eqno(28)$$
where $J$ is the Kahler form of the blown up $S^2$s.

We now show that this model has a metastable nonsupersymmetric vacuum in addition to a supersymmetric one. The F--terms derived from the superpotential are
$$F_{\Phi_1}=2m\Phi_1-Q_{12}Q_{21} \eqno(29)$$
$$F_{\Phi_2}=2m(\Phi_2-a)+Q_{12}Q_{21} \eqno(30)$$
$$F_{Q_{12}}=Q_{21}(\Phi_2-\Phi_1) \eqno(31)$$
$$F_{Q_{21}}=Q_{12}(\Phi_2-\Phi_1) \eqno(32)$$
As usual, the scalar potential is given by
$$V=|F_{\Phi_1}|^2+|F_{\Phi_2}|^2+|F_{Q_{12}}|^2+|F_{Q_{21}}|^2+g^2|D|^2 \eqno(33)$$
where $g$ is the gauge coupling (of $U(1)_1-U(1)_2$).

Minimizing the scalar potential we find that the nonsupersymmetric vacuum is at $Q_{12}=Q_{21}=\Phi_1=0$ and $\Phi_2=a$. Supersymmetry is broken since
$D=\xi \not=0$. This vacuum is locally stable only if $m_{Q_{21}}^2=a^2-2g^2 \xi>0$ so that all scalar masses squared are positive. (The other two scalar masses squared are positive 
for all values of the parameters.) We see that in this configuration one of the D5 branes is at the origin whereas the other is located at $x=a$. (Both branes are located 
at the origin of the remaining four transverse directions parametrized by $Q_{12},Q_{21}$.) In fact it is
exactly this separation between the branes that gives mass to the open strings between the D5 branes described by $Q_{12},Q_{21}$ and guarantees metastability for $a^2-2g^2 \xi>0$.
As in the previous section, supersymmetry breaking in the metastable vacuum requires a nonzero anomalous D--term; if $\xi=0$ the above vacuum becomes supersymmetric.

The superymmetric vacua are given by
$$\Phi_1=\Phi_2={a \over 2} \qquad Q_{12}={{ma} \over {Q_{21}}} \eqno(34)$$
where
$$|Q_{21}|^2={\xi \over 2} \pm {1 \over 2} \sqrt{\xi^2+4m^2a^2} \eqno(35)$$
Only the positive sign above makes sense since $|Q_{21}|^2$ cannot be negative.
As before, the VEV of $Q_{21}$ is fixed up to a phase which means that the vacuum manifold is a circle parametrized by this phase (which does not describe the moduli space). We can set this
phase to zero and thus the VEV of $Q_{12}$ is real. Then, there are two supersymmetric vacua given by eqs. (34) and (35).
In these vacua the two branes are located at the same point on $C(x)$ namely at $x=a/2$; however they are separated in the other four transverse directions. One of the branes
remains at the origin of the four transverse directions whereas the second one is located away from the origin, at a point whose coordinates are fixed by the VEVs of $Q_{12},Q_{21}$ given
by eqs. (34) and (35).

Just like the intersecting brane model of the previous section, this model is a deformation of an ${\cal N}=2$ supersymmetric model. In the present case, the relevant 
deformations are the fibering of the $A_2$ singularity over $C(x)$, the locations of the singularities and the blow up. The $x$
dependence in eq. (20) which describes the fibering gives rise to masses for $\Phi_1,\Phi_2$. The shift in the location of the second $S^2$, to $x=a$ given by the last term in eq. (21)
gives rise to masses for $Q_{12},Q_{21}$. In addition, there is the anomalous D--term which arises from the different Kahler deformations on the two blown up $S^2$s. Notice that these are
exactly the same deformations we used in section 2 for the intersecting brane construction.

We can also realize F--term supersymmetry breaking in the metastable vacuum by taking $\xi=0$, i.e. by not blowing up the singularity by the two $S^2$s. The F--terms in eqs. (29-32) and the 
scalar potential remain the same. 
As before, we satisfy the D--term constraints by taking $Q_{12}=Q_{21}=Q$. Then, the metastable nonsupersymmetric vacua are given by 
$$\Phi_1+\Phi_2=a \qquad \Phi_2-\Phi_1={{2Q^2} \over {m}} \qquad Q^2={3 \over 2}m^2 \left(-{1  \pm  \sqrt{1+{{4a} \over {9m}}}}  \right) \eqno(36)$$
As expected, all F--terms are nonzero in these vacua. We only consider the solution with the plus sign above in order to have only real VEVs.
This vacuum is locally stable if $a<16m$. Metastability arguments require that the nonsupersymmetric vacuum be parametrically far from
the supersymmetric one in terms of a small coupling. In eq. (26), the superpotential does not include a coupling constant
(the common convention in the literature) since factors of $g$ have been absorbed into the definitions of $\Phi_{1,2}$. If we restore
factors of $g$ in eq. (26) we find that the vacua are parametrically far from each other since all the VEVs in eq. (36) are
proportional to inverse powers of $g$. For small enough $g$ these vacua are metastable.

We see that in these metatstable vacua the wrapped D5 branes are separated in all transverse directions.
The supersymmetric vacuum, on the other hand, is at $Q_{12}=Q_{21}=\Phi_1=0$ and $\Phi_2=a$. This is the nonsupersymmetric vacuum for the $\xi \not =0$ case. This is not surprising 
because, as we mentioned above, if $\xi=0$ , supersymmetry cannot be broken there. (Ther is another supersymmetric vacuum at
$\Phi_1=\Phi_2=a/2$ and $Q=\sqrt{ma}$.)

Comparing the superpotentials in eqs. (17) and (26) we see that they are very similar with the identifications 
$\lambda=1$, $M=m$ and $m=a$. (The model in this section has an extra 
singlet $\Phi_1$ but this is not an important difference.) Thus the models we described in the previous and this section 
are basically the same; the former is realized by intersecting branes and the latter by wrapped branes. This becomes clearer if we make the shift $\Phi_2^{\prime}=\Phi_2-a$ and compare eqs. (17) and (26).

\bigskip
\centerline{\bf 4. Conclusions and Discussion}
\medskip

In this letter, we described two different brane constructions of models with metastable nonsupersymmetric vacua. In section 2, we constructed a model by using
intersecting branes. In section 3, we constructed a similar model by using branes wrapped on singularities. In both cases the models at the field theory level are quite simple; they 
contain an Abelian gauge group with two oppositely charged
fields and one (or two) singlets. Both types of models are obtained by deformations of ${\cal N}=2$ supersymmetric models such as mass terms for the charged and singlet fields, a tree level F--term 
and an anomalous D--term. In intersecting 
brane models these deformations are different rotations and displacements of the D4 and NS5 branes. In wrapped brane models, they correspond to fibering the singularity over a complex 
plane, the locations of the nodes and Kahler 
deformations that blow up the singularity. Using the same ideas and tools one can build a class of similar models with metastable nonsupersymmetric vacua. 

In our models, supersymmetry breaking in the metastable vacua is not dynamical since it occurs at tree level. In models with D--term supersymmetry breaking, this cannot be avoided.
In field theory, in models with F--term supersymmetry breaking, dynamical supersymmetry breaking can be realized by retrofitting[\DIN]. In string theory, one possible way to obtain dynamical 
supersymmetry breaking in the metastable vacuum is to construct models with D--brane instantons[\ABK,\INS]. In models with 
vanishing tree level singlet masses and singlet F--terms, these parameters are generated by brane instanton effects and are thus exponentially suppressed. This mechanism may lead to 
metastable vacua 
with dynamical supersymmetry breaking in a manner similar to the models above. If one of the nodes of the singularity goes through a geometric transition[\GEO], brane instanton corrections 
at that node can be calculated[\NPS]. It is harder to achieve dynamical 
supersymmetry breaking through a similar mechanism in intersecting brane models since in this case the instanton (which is an Euclidean D0-brane along the $x_6$ direction) effects are 
more difficult to calculate.

The brane models we constructed in this letter are very similar to those that realize D--term inflation[\DINF]. The only difference is the tree level masses in the superpotential which vanish
in D--term inflation models. However, these need not vanish for inflation to occur. Since the singlet is the inflaton in these models its mass has to be smaller than the Hubble
constant for slow--roll inflation to take place. Moreover, the mass for one of the charged fields has to be small enough to turn the origin into an unstable state with a tachyonic direction
in field space. This field plays the role of the trigger field in D--term inflation.
We see that the conditions for inflation are somewhat complementary to those for a metastable vacuum with broken supersymmetry since in the former we need the origin to be unstable whereas
in the latter it has to be metastable. Nevertheless, the same models we constructed above (with small masses) can be used for realizing D--term inflation in string theory[\EDI].
This requires that they be compactified which rules out intersecting brane models but not those with branes wrapped on singularities.


\bigskip
\centerline{\bf Acknowledgements}

I would like to thank the Stanford Institute for Theoretical Physics for hospitality.

\vfill

\refout

\end
\bye